# Intermittent synchronization in a network of bursting neurons

Choongseok Park[1] (박중석), Leonid L. Rubchinsky[1,2]

[1] Department of Mathematical Sciences and Center for Mathematical Biosciences, Indiana University Purdue University Indianapolis, Indianapolis, IN 46202, USA
[2] Stark Neurosciences Research Institute, Indiana University School of Medicine, Indianapolis, IN 46202, USA

## Abstract

Synchronized oscillations in networks of inhibitory and excitatory coupled bursting neurons are common in a variety of neural systems from central pattern generators to human brain circuits. One example of the latter is the subcortical network of the basal ganglia, formed by excitatory and inhibitory bursters of the subthalamic nucleus and globus pallidus, involved in motor control and affected in Parkinson's disease. Recent experiments have demonstrated the intermittent nature of the phase-locking of neural activity in this network. Here we explore one potential mechanism to explain the intermittent phase-locking in a network. We simplify the network to obtain a model of two inhibitory coupled elements and explore its dynamics. We used geometric analysis and singular perturbation methods for dynamical systems to reduce the full model to a simpler set of equations. Mathematical analysis was completed using three slow variables with two different time scales. Intermittently synchronous oscillations are generated by overlapped spiking which crucially depends on the geometry of the slow phase plane and the interplay between slow variables as well as the strength of synapses. Two slow variables are responsible for the generation of activity patterns with overlapped spiking and the other slower variable enhances the robustness of an irregular and intermittent activity pattern. While the analyzed network and the explored mechanism of intermittent synchrony appear to be quite generic, the results of this analysis can be used to trace particular values of biophysical parameters (synaptic strength and parameters of calcium dynamics), which are known to be impacted in Parkinson's disease.

**Synchronized neural oscillations are widespread phenomena with a variety of functional implications and have been observed in many neural systems. In particular, synchronized oscillations of bursting neurons are related to motor symptoms of Parkinson's disease. Recent experiments revealed intermittent characteristic of this synchronized activity, which may have functional importance. This manuscript explores potential mechanisms underlying this intermittent synchronization. We reduced large network model [1] to a simpler model of two coupled bursting neurons and used fast/slow analysis to explore the mechanisms of intermittent synchronization. Intermittently synchronous oscillations are generated by overlapped spiking which crucially depends on the geometry of slow phase**

plane and the interaction between slow variables as well as the strength of coupling between bursting cells.

## 1. Introduction

Synchronized oscillations across various brain areas have been extensively studied because of their functional significance for perceptive and cognitive processing [2,3,4] and for movement preparation and execution [5,6,7]. Thus it is natural that disorganization of this oscillatory activity (including pathologically strong or pathologically weak synchronization of oscillations) may contribute to a variety of neurological and psychiatric disorders [8,9]. In particular, hypokinetic motor symptoms of Parkinson's disease such as slowness and rigidity of voluntary movements are closely related to synchronized oscillatory neuronal activity in the beta frequency band, loosely speaking, 10-30 Hz [10,11,12]. This kind of synchronization is intermittent and has a specific temporal patterning [13].

Earlier modeling and experimental studies investigated the role of two brain nuclei – external globus pallidus (GPe) and subthalamic nucleus (STN) in the support of this oscillatory synchronized dynamics in Parkinson's disease. In particular, synchronized oscillations have been suggested to result from rhythmic sequences of recurrent excitation and inhibition in pallido-subthalamic networks [14,15]. While the role of other structures in the generation of these synchronized oscillations cannot be ruled out, our own modeling work suggested that the characteristic temporal structure of oscillatory activity in the beta-band may be due to intrinsic properties of the STN-GPe network itself and showed that the corresponding parameter region resides between synchronized and nonsynchronized oscillatory dynamics [1].

The transient nature of neural dynamics should not be surprising [16,17]. However, although many network architectures and their activity patterns have been extensively studied, biological and dynamical mechanisms underlying synchrony with irregular desynchronizing events (like those described in [13]) are not fully understood. In the current study, we investigated one possible generic mechanism underlying this intermittent synchrony in networks of inhibitory and excitatory bursting neurons. We reduced the excitatory-inhibitory network formed by STN and GPe to a simple network consisting of two inhibitory cells with self-inhibition. The resulting simple model is not able to fully capture all the details of the complex network activity patterns shown in experimental recordings [13] and previous modeling work [1]. However, it captures important characteristics of intermittent synchrony and provides insights into potential dynamical and biological mechanisms and their dependence on parameters.

We used geometric dynamical systems and singular perturbation methods to reduce the full model to a simpler set of equations. This approach simplifies the analysis of the dynamics and investigation of the underlying mechanisms. Mathematical analysis was completed using three slow variables with two different time scales. We show how two slow variables are responsible for the competition between two cells by studying the geometry of slow phase plane. This competition is responsible for the generation of overlapped spiking. The slowest variable, on the other hand, broadens the range of synaptic strength, over which irregular and intermittent activity patterns occur. The results in this study describe how the properties of the cells and strength of synaptic connections interact to generate intermittent synchrony and may be used to relate the model parameter relationship to the properties of real systems.

In Section 2, a conductance-based model of the STN cell and network architecture is introduced. In Section 3, we study various activity patterns focusing on the intermittent phase-locking due to overlapped spiking of neurons. In Section 4, we provide a geometric analysis of intermittent synchronization using fast/slow analysis. Overlapped spiking and regular bursting solutions are subjected to the analysis based on the geometry of a two-dimensional bifurcation diagram.

## 2. Neural Model and Network Architecture

Experimental results and previous modeling studies have suggested that the STN-GPe network within the basal ganglia may form a key network and their interactions play a crucial role in the pathophysiology of Parkinson's disease [14,15]. In the reciprocally connected STN-GPe network, the excitatory input from STN to GPe is relatively sparse whereas the inhibitory input from GPe to STN is relatively dense [18]. The GPe cell also gets an external inhibitory input from striatum. Previous modeling works [1,19,20] considered a network consisting of 10 STN cells and 10 GPe cells where STN cell sends an excitatory input to a GPe cell and GPe cell sends an inhibitory input to nearby STN cells (Fig.1A). Such networks are able to produce realistically rich patterns of activity. In this study, we will start with this network architecture and reduce it to a simpler one as follows.

The aforementioned modeling studies have shown that a transition from an irregular, uncorrelated regime (presumably close to the normal state) to a regular, synchronized regime (presumably extreme pathology) in STN-GPe network can be accomplished in a biologically realistic manner by controlling two dopamine-dependent parameters, more hyperpolarizing current input to GPe and stronger inhibitory synaptic connection from GPe to STN. Under more hyperpolarizing current, the model GPe cell becomes less active and spontaneous activity is substantially reduced. Thus, as we approach a synchronized regime through intermittent synchrony, the GPe cell tends to relay the excitation from the STN cell faithfully due to its reduced spontaneous firings. For simplicity of analysis, we assume that the GPe cell acts as a simple relay cell and reduce the whole network to a network of STN cells. Under this assumption, activity of the STN cell is instantaneously delivered to a GPe cell and that GPe cell faithfully reproduces STN cell activity, hence the STN cell and the GPe cell, which gets an excitatory input from that STN cell, are merged together. Thus although real STN cells are excitatory, the resulting network consists of cells with inhibitory connections. In particular, the reciprocal inhibition from a GPe cell to an STN cell becomes self-inhibition. Therefore, we considered a network consisting of two inhibitory cells with reciprocal inhibition as well as self-inhibition (Fig. 1B). We assumed that the strength of self-inhibition is 30% of the reciprocal synaptic strength unless specified otherwise.

We employ the conductance-based single compartment model of STN cells used in [1,19,20], which consists of membrane potential (*V*), three gating variables for ionic currents (*n*, *h* and *r*), intracellular concentration of $Ca^{2+}$, and synaptic variable (*s*). The equations are

$$C \cdot dV/dt = -I_L - I_K - I_{Na} - I_T - I_{Ca} - I_{AHP} - I_{syn} + I_{app} \quad (1)$$

$$dx/dt = \phi_x(x_\infty(V) - x)/\tau_x(V) \text{ for } x = n, h, r \quad (2)$$

$$d[Ca]/dt = \varepsilon(-I_{Ca} - I_T - k_{Ca}[Ca]) \quad (3)$$

$$ds/dt = \alpha H_\infty(V_{presyn} - \theta_g)(1-s) - \beta s \quad (4)$$

$I_L = g_L(V - V_L)$, $I_K = g_K n^4 (V - V_K)$, $I_{Na} = g_{Na} m_\infty^3(V) h(V - V_{Na})$ are leak, fast potassium and sodium currents, respectively. $I_T = g_T a_\infty^3(V) b_\infty^2(r)(V - V_{Ca})$ and $I_{Ca} = g_{Ca} s_\infty^2(V)(V - V_{Ca})$ are two calcium currents and $I_{AHP} = g_{AHP}([Ca]/([Ca]+k_1))(V - V_K)$ is an afterhyperpolarization current. $m_\infty$, $a_\infty$, and $s_\infty$ are instantaneous voltage-dependent gating variables. $I_{syn} = g_{syn}(V - V_{syn}) \sum_j s_j$ is synaptic current where summation is over *s*-variables from all neurons projecting to a given neuron. Sigmoidal function $H_\infty$ is given by $1/(1+\exp[-(V - \Theta_g^H)/\sigma_g^H])$. The parameters are the same as in [20] with the minor exception of $g_T$ changed from 0.5 to 0.6.

In the later part of this study, we will focus on the dynamics of three slow variables (the gating variable *r* of $I_T$, total synaptic input σ, and the calcium concentration [*Ca*]) with their time scales. The total synaptic input for cell 1, say $\sigma_1$, is defined as $g_{syn}(0.3 s_1 + s_2)$ where $s_1$ is the synaptic variable of self-inhibition and $s_2$ is the synaptic variable for synapse from the other cell. The time constant $\tau_r(V)/\phi_r$ for *r* in Eq. (2) is 49.2 over most of voltage range, 1/β for *s* in Eq. (4) is 25, and 1/ε for [*Ca*] in Eq.(3) is 2667 while membrane potential *V* rises and falls within few milliseconds when a cell fires. We have two slow variables σ and *r*, and another extremely slow variable [*Ca*].

This model was integrated using XPPAUT. Customized MATLAB codes were used for time-series analysis which is described in more detail in [13,21]. Delays due to conductance, synapses and cell dynamics may influence the intermittent synchronization properties observed in experimental data. However, this simple network is able to generate patterns, which are qualitatively similar to those observed in experimental data. Since we are interested in the generic mechanism underlying such activity patterns, this network architecture is justified for our study.

The resulting network appears to be relatively typical for neural systems. Its mutually inhibitory organization makes it prone to the generation of rhythms, which makes it somewhat similar to central pattern generator (CPG) networks.

# 3. Synchronous activity patterns in the network

## 3.1 Synchronous activity in dependence on the coupling strength

We chose $g_{syn}$ (inhibitory synaptic strength) as a varying parameter (it is one of the parameters, expected to increase as the dopamine disappears in Parkinson's disease [22,23,24,25]) and first consider activity patterns over various $g_{syn}$ values. Without self-coupling, we observed out-of-phase bursting solutions (phases of bursting in neurons are shifted by $\pi$) for sufficiently large $g_{syn}$ values due to the post-inhibitory rebound burst of STN cell [20]. With self-coupling, we also observe out-of-phase bursting solutions where the number of spikes within a burst of the out-of-phase solution increases monotonically as $g_{syn}$ increases. But, there are also intervals of parameters, which correspond to transitions from one out-of-phase solution to another. Over such intervals, activity patterns demonstrate a mixture of the two nearby solutions. The lengths of such intervals become shorter as $g_{syn}$ increases.

Note that while [20] studied similar models, it looked at extremes: clustered regime in the strong coupling case and irregular regime in the weak coupling case. This study looks at the intermediate case, when the coupling is intermediate and the dynamics is neither perfectly clustered nor irregular. Not only this dynamics is more complex, it is also more realistic, as we show below. The present study outlines a scenario for a realistically complex dynamics, advancing us beyond well-studied cases of weak and strong coupling.

To characterize the dynamics of the network, we consider the correlation between activities of the two cells in a large range of coupling strength. First, we computed coherence in the frequency domain using membrane potentials. We downsampled 10s-long voltage data to 1 kHz and divided it into 8 sections with 50% overlap. A Welch's averaged, modified periodogram method was used to compute the coherence estimate over frequencies. Fig. 2A shows the coherence values averaged over [0,100] Hz (black) and [10, 30] Hz (gray) frequency bands. We observe an overall increase of average coherence in both cases as $g_{syn}$ increases because network activity patterns are controlled by stronger inhibitory connection.

This network can generate complex activity patterns over a significant interval of parameter values due to reciprocal inhibitory input. One example is for $g_{syn}$ around 0.9 (see Fig. 2A), where the average coherence over 10-30Hz frequency band shows a dip. Interestingly, solutions over this interval demonstrate substantial power in 10-30 Hz frequency band. Fig. 2C illustrates activity patterns when $g_{syn} = 0.9$. This solution "lies between" 2-spike and 3-spike out-of-phase bursting solutions and shows alternation of bursts, the number of spikes within each cycle tends to change irregularly. To quantify this imperfect synchrony, we computed the degree of phase-locking between two cells using the synchronization index $\gamma$. We first band-pass filtered the signal within 10-30 Hz band and constructed phase variables using Hilbert transform. The synchronization index is defined as $\gamma_N(t_k) = \left\| \frac{1}{N} \sum_{j=k-N+1}^{k} e^{i\phi_j} \right\|^2$, where $\phi_j$ is the difference of phases and $N$ is the number of data points over a window of some fixed length. We used 1s long non-overlapping windows and averaged the resulting synchronization indices to get an averaged synchronization index. Fig. 2B shows the averaged synchronization index over some $g_{syn}$ values and different self-coupling strengths. When the self-coupling strength is 30%, we have 1 or 2-spike (3-spike) out-of-phase bursting solution for $g_{syn}$ smaller than 0.9 (greater than 0.93). Over these two ranges of $g_{syn}$ values, we have relatively strong phase locking between the two cells. This secure phase locking is reduced when $g_{syn}$ is between 0.9 and 0.92, where an irregular sequence of burstings with overlapped spikes dominates. Different values of self-coupling preserve this characteristic feature of phase-locking over a range of $g_{syn}$ values while the $g_{syn}$ value which shows the most significant drop of phase-locking decreases as the strength of self-coupling increases.

## 3.2 Irregular Dynamics

Now let us explore whether the activity pattern shown in Fig. 2B is really chaotic. To do it numerically we compute maximal Lyapunov exponents (MLEs); in general, a positive maximal Lyapunov exponent indicates exponential growth of small perturbations and chaotic dynamics if the solutions are bounded. Instead of using whole time series, we may use one-dimensional map data obtained through Poincare section. We chose the gating variable $r$ of $I_T$ (Eq. (1)) in one of the cells for Lyapunov exponent computation. $I_T$ is essential for the generation of a rebound burst after removal of prolonged hyperpolarization, which, in turn, is important for generating irregular activity patterns. We recorded values of $r$ of STN 1 whenever it is released from inhibition to get $\{r_n\}$. We say that STN 1 is released from inhibition when total synaptic input $\sigma_1$ of STN 1 begins to decrease after STN 2 fires its last spike of a burst. Fig. 3A shows return maps, $r_{n+1}$ vs. $r_n$, for several different $g_{syn}$ values from 0.86 through 0.96. For $g_{syn} \leq 0.89$ or $g_{syn} \geq 0.93$, the return map has only one single point; black square (circle, triangle, diamond) for $g_{syn} = 0.86$ (0.88, 0.94, 0.96, respectively). A single point (a tight cluster of points) on $r_{n+1} = r_n$ line implies an almost regular and periodic solution. In this case the black square and circle are 2-spike bursting solutions, the triangle and diamond are 3-spike bursting solutions. Between these two cases, we have more complicated return maps when $g_{syn}=0.9$ (Fig. 3B) and 0.92 (Fig. 3A, gray dots). The complex appearance of the return map in Fig. 3B may suggest chaotic dynamics. And indeed, the maximal Lyapunov exponent is 0.5292 when $g_{syn} = 0.9$, hence activity patterns are really irregular. $g_{syn} = 0.92$ case (Fig. 3A) also has positive maximal Lyapunov exponent 0.3613. Fig. 3C shows the maximal Lyapunov exponents over other $g_{syn}$ values from 0.894 to 0.924. The values of maximal Lyapunov exponents show complex dependence on $g_{syn}$ and are relatively large for $g_{syn}$ between 0.9 and 0.91. Maximal Lyapunov exponents were computed with TISEAN package [26].

### 3.3 Escaping and intermittent phase synchronization

The characteristic irregular sequence of burstings shown in Fig. 2B may be attributed to the competition between the two cells, which is evidenced by frequent occurrences of overlapped spikes. This kind of spike overlapping can be characterized as 'escape' [27]. The observed variability of number of spikes associated with an escape mechanism is responsible for the reduction of the phase-locking strength. Fig. 4A shows an example of 'escape' in detail when $g_{syn} = 0.9$. Two voltage profiles are shown in upper panel where gray trace is escaping from inhibition of the other cell and the corresponding phase variables are shown in bottom panel. We can see that overlapped spikes cause distortions of the phase relationship between the two cells through delay and/or advance of phase propagations, which would be almost constant in case of regular bursting solutions. The competition of the two cells through escaping redefines the relationship between them. The accumulated effect of this competition is shown in Fig. 4B through the first-return map for the phase difference. We recorded the phases of cell 1 whenever the phase of cell 2 passes upward through some fixed value. In this figure, we set the check point as -3 for cell 2 and collected $\{\phi_i\}$ for cell 1. $\{\phi_i\}$ was shifted so that the mean is $\pi/2$ and then we plotted $\phi_{i+1}$ vs. $\phi_i$.

In Fig. 4B, one big cluster in the first quadrant corresponds to the synchronized activity patterns between two cells, while two big branches in the second and fourth quadrants correspond to desynchronized activity. Deviation from the cluster is mostly due to the frequent occurrence of overlapped spikes and is characterized by a quick return to the cluster (Fig. 4C). Similar first-return maps for the phase difference and relatively short duration of desynchronizing events was also observed in experimental data in parkinsonian patients [13] and in the model of the large basal ganglia network [1].

Although the origin and the underlying mechanisms of this spike overlapping and resulting intermittent synchronization will be discussed in the next part of this paper, one heuristic explanation can be given. Irregular and frequent occurrence of overlapping is due to the specific nature of the given solution itself; in the parameter space it is situated in between two stable solution regimes. The solution tends to be either in 2-spike or 3-spike bursting modes but this tendency for stability is frequently destroyed by other destabilizing factors. We would like to stress that this activity pattern is robust in numerical simulations with weak additive Gaussian white noise, which suggests that this mechanism of activity pattern may be of experimental relevance.

# 4. Geometric analysis of intermittent synchronization

## 4.1 Slow phase plane construction via fast/slow analysis

In this section, we will investigate how frequent occurrences of overlapped spiking, which result in characteristic irregular sequence of burstings and intermittent phase synchronization, depend on the strength of synaptic connection, intrinsic properties of the cells and the interplay between slow variables. We will use geometric methods and begin by considering a STN cell when $g_{syn}$ = 0.9 (intermittent phase-locking case illustrated in Figs. 3 and 4). The voltage profile is shown in the upper panel of Fig. 5A and the lower panel shows three other variables, the gating variable $r$ of $I_T$, the total synaptic input $\sigma$, and the calcium concentration $[Ca]$. Total synaptic input $\sigma$ is the sum of inhibitory input from the other cell and self-inhibition.

Due to the slowest time scale of $[Ca]$, we may consider it as a constant and proceed with fast/slow analysis. It seems, however, that $[Ca]$ plays an important role in the generation of irregular sequence of burstings with overlapped spiking. We inspected activity patterns under constant $[Ca]$ ranged from 0.66 to 0.70 and found that they tend to be periodic. These periodic solutions are not like simple out-of-phase 2-spike or 3-spike solutions but more complicated patterns with escaping. For example, when $[Ca]$ =0.67 and $g_{syn}$ = 0.9, one period of activity patterns consists of 1) 3-spikes burst beginning with escaping, 2) one alternation of bursts each consisting of 3-spikes and 3) 3-spikes burst ending with escaping. In some case, we have chaotic-like activity patterns also. The activity patterns when $[Ca]$ = 0.68 and $g_{syn}$ =0.9 provides one example. In this case, long repetition of periodic patterns stated above is interrupted by intermittent occurrences of other types of long burst alternation. For other values of $g_{syn}$, however, periodic patterns appear again. As a result, the range of $g_{syn}$ values over which chaotic patterns appear is significantly reduced. In other words, the occurrence of irregular patterns shows sensitivity to $[Ca]$ values. Hence we include $[Ca]$ in our fast/slow analysis.

To begin fast/slow analysis, we first regard $[Ca]$ as a constant ($[Ca]$ = 0.7 in the following analysis unless mentioned otherwise), while $\sigma$ and $r$ are slow variables. Then we can explore the bifurcation diagram of the fast subsystem with parameters $\sigma$ and $r$. We first fix $r$ and consider $\sigma$ to be a bifurcation parameter. Fig. 5B shows the resulting bifurcation diagram when $r = 1$. We have Z-shaped curve of fixed points. For larger values of $\sigma$, there are three fixed points; the lower fixed point is stable, the middle is a saddle, and the upper is unstable. As $\sigma$ decreases, lower stable and middle saddle fixed points merge at a saddle-node bifurcation (labeled SN). There is also a subcritical Hopf bifurcation point on the upper branch and fixed points become stable once passed this point (thick black). A branch of the unstable periodic orbit (thin gray) which becomes to be stable (thick black) emanates from the Hopf bifurcation point and becomes a saddle-node homoclinic orbit when $\sigma = \sigma_{SN}$. In fact, this bifurcation structure persists for each $r$ on [0, 1].

We trace the saddle-node bifurcation point (SN) in the bifurcation diagram as $r$ varies to get a two dimensional bifurcation diagram which is shown in Fig. 6A. We call the resulting curve Σ-curve (gray curve in the (σ, r) plane at Fig. 6A). The fast subsystem shows sustained spiking in the region left to Σ (spiking region) and quiescence in the region right Σ (silent region). Note that if r is sufficiently small, then we cannot get an oscillatory solution. Fig. 6A also shows frequency curves (dependence of frequency of spikes on the total synaptic input σ for different values of r) in the spiking region. Fig. 6B provides another view of these curves. There is a band-like region of lower frequency along Σ, visible in the frequency curve when r = 0.25. This band is more prominent along the lower part of Σ and this will play an important role in the generation of overlapped spiking.

## 4.2. Regular out-of-phase bursting solutions in the phase plane of slow variables and linear stability under constant calcium level

Fig. 7 shows the two parameter bifurcation diagram with the projection of regular 2-spike out-of-phase bursting solution when $g_{syn}$ = 0.86. Without loss of generality, let's assume that active cell is cell 2 and silent cell is cell 1. We will follow trajectories of both cells from the moment when cell 2 fires its second spike. Upper filled circle in Fig. 7 denotes ($\sigma_1$, $r_1$) of cell 1 and lower filled circle denotes ($\sigma_2$, $r_2$) of cell 2 at this moment.

First note that synaptic variable s of a cell rises once membrane potential rises, passes certain threshold ($\theta_g$), and stays above it; s decreases otherwise (Eq. 4). More precisely, if membrane potential is larger than $\theta_g$, then s quickly approaches $\alpha/(\alpha+\beta)$, and if it is less than $\theta_g$, then s slowly approaches 0 (Eq. 4). Hence s repeats ups and downs when a cell fires a series of action potentials, while it decreases monotonically when a cell is silent. We also note that total synaptic input σ is a weighted sum of synaptic variables multiplied by synaptic strength. For example, $\sigma_1$ = $g_{syn}$ (0.3* $s_1$+ $s_2$). Therefore, even though a cell is in active phase, corresponding total synaptic input σ shows wiggles with reduced amplitude due to self-inhibition (Fig. 5A).

Now let's return to the trajectories in Fig. 7. When cell 2 fires its second spike, $s_2$ approaches $\alpha/(\alpha+\beta)$ quickly but $s_1$ is small (around 0.1) and keeps decreasing, hence the increment of $\sigma_1$ is near $g_{syn}\,\alpha/(\alpha+\beta)$ while the increment of $\sigma_2$ is near 30% of $g_{syn}\,\alpha/(\alpha+\beta)$. Due to this large increment in $\sigma_1$, ($\sigma_1$, $r_1$) is shifted to the silent phase (open circle) while ($\sigma_2$, $r_2$) is slightly shifted to the right. Once membrane potential drops below $\theta_g$, both $s_1$ and $s_2$ begin to decrease and so do both $\sigma_1$ and $\sigma_2$.

To understand dynamics of r when cell 2 fires its second spike, we need to consider the governing equation of r (Eq. 2). For the clarity of explanation, we rewrite it here:

$$\frac{dr}{dt} = \phi_r \frac{r_\infty(V)-r}{\tau_r(V)}$$

The dynamics of $r$ is driven by two factors: $r_\infty(V)$ and the time constant $\tau_r(V)/\phi_r$. For example, suppose that membrane potential $V$ is fixed at some value $V_0$. Then $r$ approaches $r_\infty(V_0)$ over time. How fast or slowly $r$ approaches $r_\infty(V_0)$ is determined by the time constant $\tau_r(V_0)/\phi_r$. The smaller the time constant is, the faster $r$ approaches $r_\infty(V_0)$. As was stated before, $\tau_r(V)/\phi_r$ is 49.2 over reasonable range of voltage $V$. Thus, the dynamics of $r$ is mainly driven by the values of $r_\infty(V)$. Value of $r_\infty(V)$ is close to zero if membrane potential $V$ is sufficiently large and is close to one for sufficiently small values of $V$. There is a monotonic transition between these two states over intermediate values of $V$. When a cell fires an action potential, membrane potential goes through three stages: 1) fast up and down of voltage, 2) afterhyperpolarization period, and 3) recovery period. Accordingly, $r_\infty(V)$ takes small values close to zero during up/down strokes, large values close to one during afterhyperpolarization period, and then intermediate values over recovery period. This explains the behavior of $r_2$ in Fig.7, where $r_2$ decreases initially, increases for a while, and then decreases again. The dynamics of $r_1$ is more simple. When cell 2 fires its second spike, $r_1$ first increases due to the hyperpolarization caused by inhibitory input from cell 2 and then decreases as the inhibition wears off.

Let us consider how far $r_1$ can go under hyperpolarization. It is an important issue in the current study because larger value of $r$ may result in longer duration of post-inhibitory rebound burst [20,28] when a cell is released from inhibition and enters spiking region (Fig. 8). The increment of $r_1$ under hyperpolarization depends on the following two factors: degree of hyperpolarization and duration of hyperpolarization. Degree of hyperpolarization is a function of synaptic strength ($g_{syn}$) because $g_{syn}$ determines how much the membrane potential $V$ can be lowered by inhibitory input from the other cell. Admitting that the dynamics of $r$ is driven by $r_\infty(V)$ and $r_\infty(V)$ takes values close to 1 over sufficiently small $V$, $g_{syn}$ controls the maximum level that $r$ may reach during hyperpolarization. In other words, the degree of hyperpolarization which is modulated by $g_{syn}$ controls the maximum level of $r_1$. On the other hand. if the duration of hyperpolarization is not sufficient, then $r_1$ may not reach the maximum set up by $g_{syn}$.

After some time past Σ in the (σ, $r$) plane, cell 1 finally fires its first spike (upper square, Fig.7). We denote the time needed for cell 1 to fire its first spike after it is released from inhibition (from upper open circle to upper square in Fig. 7) by $T_1$. It is not clear here whether cell 2 were able to fire another spike or not if it were not interrupted by cell 1. If cell 2 were able to fire another spike, then the inter-spike interval between its second and third spikes should be greater than $T_1$ of cell 1. If not, then cell 2 would get a chance to fire its third spike and cell 1 be inhibited again. In fact, the lower frequency band along Σ make this scenario possible because cell 2 traverses near the lower frequency band along Σ after its second spike and this results in longer inter-spike interval between second and third spikes of cell 2. It is also possible that cell 2 simply leaves the active region after its second spike.

Once cell 1 fires its first spike, due to the self-inhibition, ($\sigma_1$, $r_1$) is also pushed to the right very quickly and then begins to move to the left slowly to enter the spiking region again to fire its next spike, while ($\sigma_2$, $r_2$) traverses mostly silent region. Because the inter-spike interval between first and second spikes of cell 1 is smaller than the time needed for cell 2 to enter the spiking region and fire its first spike, cell 1 fires its second spike. This comprises the half cycle of the regular 2-spike bursting solution in the ($\sigma$, $r$) plane of slow variables.

To get some insight into the stability of 2-spike regular out-of-phase bursting solution, we simplify the dynamics of $r$ by averaging its governing equation. Recall that overall level of $r$ is increasing in silent region and decreasing in spiking region although it shows up and down fluctuations. Let $r_\infty^S$ ($r_\infty^A$) be the averaged value of $r_\infty(V)$ over silent (active) phase in the governing equation of $r$ (Eq. 2). Here we say that a cell is in the silent phase if it is in the silent region (right of the $\Sigma$) and is in the active phase if it is in the spiking region. We also assume that $\tau_r(V)$ is constant, say $\tau_r$, because $\tau_r(V)$ is almost constant over a physiologically relevant range of $V$. Then the averaged equation is given by $r' = k(\bar{r}_\infty - r)$ where $k = \phi_r / \tau_r$ and $\bar{r}_\infty$ is either $r_\infty^S$ or $r_\infty^A$ with general solution $r = \bar{r}_\infty + (r_0 - \bar{r}_\infty)e^{-kt}$ for some initial condition $r_0$. In this averaged framework $r$ is monotonically increasing (decreasing) over silent (active) phase.

We define $T_I$ as the time from the point when silent cell is released from inhibition to the point when ($\sigma$, $r$) reaches $\Sigma$ and enters spiking region. We also define $T_A$ as the time from $\Sigma$ to its second spike. As compared to $T_1$ and $T_2$ in Fig. 7, $T_I$ is smaller than $T_1$ and $T_A$ is bigger than $T_2$. We may assume that silent cell fires its first spike when ($\sigma$, $r$) reaches $\Sigma$ and enters spiking region as in [20]. Now, it takes $T_I$ for silent cell to enter spiking region and fire its first spike. Then silent cell becomes active cell and it takes $T_A$ to fire its second spike. After its second spike, active cell spends $T_I$ in the spiking region before it gets inhibition from the other cell. Now active cell becomes silent cell and it spends $T_A$ until it is released from inhibition.

$T_I$ and $T_A$ are functions of the values of ($\sigma$, $r$) when silent cell is released from inhibition (e.g. at the open circle in Fig. 7). Recall that synaptic variable of active cell approaches $\alpha/(\alpha+\beta)$ quickly when it fires its second spike while synaptic variable of silent cell is small and keeps decreasing. Near the regular 2-spike out-of-phase solution, the values of synaptic variable of active cell are similar in magnitude. In addition, synaptic variable of silent cell is multiplied by $0.3 * g_{syn}$ where $g_{syn}$ is 0.86. Thus $\sigma$ of silent cell when it is released from inhibition is dominated by $g_{syn}\alpha/(\alpha+\beta)$. In fact it takes almost identical values with negligible differences near the regular 2-spike out-of-phase solution. By letting $\sigma$ being fixed when silent cell is released from inhibition, we may regard $T_I$ and $T_A$ as functions of $r$.

We will now trace values of $r$ when silent cell is released from inhibition and construct a map for it. If fixed point of this map is stable, then we may conclude that 2-spike regular out-of-phase bursting solution is also stable. Let $r_*^u$ be the value of $r$ when silent cell is released from inhibition and $r_*^l$ the value of $r$ when active cell fires its second spike. Now we trace $r_1$ from $r_*^u$ along the simplified trajectory of regular 2-spike bursting solution. When cell 1 fires its first spike, $r_1$ is given by $r_1^1 = r_\infty^S + (r_* - r_\infty^S)e^{-kT_I(r_*^u)}$. In the active region, it fires its second spike after $T_A(r_*^u)$ and spends $T_I(r_*^u)$ until it gets inhibition. At this time, $r_1$ is $r_1^2 = r_\infty^A + (r_1^1 - r_\infty^A)e^{-k(T_A+T_I)}$. After $T_I$, $r_1$ becomes $r_*^u$, where $r_*^u = r_\infty^S + (r_1^2 - r_\infty^S)e^{-kT_I}$. Combining these three equations, we have an implicit equation for $r_*^u$

$$r_*^u = \frac{C_3 + C_2 e^{-kT_A} + C_1 e^{-k(T_I+2T_A)}}{1 - e^{-2k(T_I+T_A)}} \tag{5}$$

where $C_1 = r_\infty^S(1-e^{-kT_I})$, $C_2 = r_\infty^A(1-e^{-k(T_I+T_A)})$, and $C_3 = r_\infty^S(1-e^{-kT_A})$. Now let $r_1^0$ ($r_2^0$) be a value of $r_1$ ($r_2$) near $r_*^u$ ($r_*^l$) when cell 2 fires its second spike and we trace them in the slow phase plane. Note that $T_I$ and $T_A$ are functions of $r_1^0$ during the first half cycle. After $T_I(r_1^0)$, we have

$$\begin{pmatrix} r_u^0 \\ r_l^0 \end{pmatrix} \mapsto \begin{pmatrix} r_\infty^S + (r_u^0 - r_\infty^S)e^{-kT_I} \\ r_\infty^A + (r_l^0 - r_\infty^A)e^{-kT_I} \end{pmatrix} \stackrel{let}{=} \begin{pmatrix} r_u^1 \\ r_l^1 \end{pmatrix} \tag{6}$$

Again, after $T_A(r_1^0)$, we have

$$\begin{pmatrix} r_u^1 \\ r_l^1 \end{pmatrix} \mapsto \begin{pmatrix} r_\infty^A + \{r_\infty^S + (r_u^0 - r_\infty^S)e^{-kT_I} - r_\infty^A\}e^{-kT_A} \\ r_\infty^S + \{r_\infty^A + (r_l^0 - r_\infty^A)e^{-kT_I} - r_\infty^S\}e^{-kT_A} \end{pmatrix} \stackrel{let}{=} \begin{pmatrix} r_u^2 \\ r_l^2 \end{pmatrix} \tag{7}$$

We let

$$F_1(x) = r_\infty^A + \{r_\infty^S + (x - r_\infty^S)e^{-kT_I(x)} - r_\infty^A\}e^{-kT_A(x)}$$
$$F_2(x,y) = r_\infty^S + \{r_\infty^A + (y - r_\infty^A)e^{-kT_I(x)} - r_\infty^S\}e^{-kT_A(x)} \tag{8}$$

Then Eq. (7) becomes

$$\begin{pmatrix} r_u^2 \\ r_l^2 \end{pmatrix} = \begin{pmatrix} F_1(r_u^0) \\ F_2(r_u^0, r_l^0) \end{pmatrix} \tag{9}$$

Similarly, over the remaining half cycle,

$$\begin{pmatrix} r_l^2 \\ r_u^2 \end{pmatrix} \mapsto \begin{pmatrix} F_1(r_l^2) \\ F_2(r_l^2, r_u^2) \end{pmatrix} = \begin{pmatrix} F_1(F_2(r_u^0, r_l^0)) \\ F_2(F_2(r_u^0, r_l^0), F_1(r_u^0)) \end{pmatrix} \tag{10}$$

where $T_I$ and $T_A$ are functions of $r_l^2$.

Combining these, we have the following map

$$\begin{pmatrix} r_u^0 \\ r_l^0 \end{pmatrix} \mapsto \begin{pmatrix} F_2(F_2(r_u^0, r_l^0), F_1(r_u^0)) \\ F_1(F_2(r_u^0, r_l^0)) \end{pmatrix} \tag{11}$$

If we linearize this map at ($r_*^u$, $r_*^l$), we have the following Jacobian matrix

$$\begin{bmatrix} \dfrac{\partial F_2}{\partial x} & \dfrac{\partial F_2}{\partial y} \\ \dfrac{\partial F_1}{\partial x} & \dfrac{\partial F_1}{\partial x} \end{bmatrix}^2 \tag{12}$$

where

$$\frac{\partial F_1}{\partial x} = (r_\infty^S - r_\infty^A) e^{-kT_A(r_*^u)}(-kT_A'(r_*^u)) + e^{-k(T_A(r_*^u) + T_I(r_*^u))} + (r_*^u - r_\infty^S) e^{-k(T_A(r_*^u) + T_I(r_*^u))}(-k(T_I'(r_*^u) + T_A'(r_*^u)))$$

$$\frac{\partial F_1}{\partial y} = 0$$

$$\frac{\partial F_2}{\partial x} = (r_\infty^A - r_\infty^S) e^{-kT_A(r_*^u)}(-kT_A'(r_*^u)) + (r_*^l - r_\infty^A) e^{-k(T_A(r_*^u) + T_I(r_*^u))}(-k(T_I'(r_*^u) + T_A'(r_*^u)))$$

$$\frac{\partial F_2}{\partial y} = e^{-k(T_A(r_*^u) + T_I(r_*^u))}$$

(13)

In the current example, we have $r_\infty^A = 0.3106$, $r_\infty^S = 0.6606$, $r_*^u = 0.45$, $r_*^l = 0.39$, $T_I = 13$ and $T_A = 41$. Numerically computed $dT_I/dr(r_*^u) = -16$ and $dT_A/dr(r_*^u) = -27$. Using these values, we found the eigenvalues of linearized map as 0.142 and 0.099, which implies that 2-spike solution is stable.

## 4.3. Dependence of number of spikes in a burst on the slow variables *r* and [*Ca*]

To analyze the dynamics that underlies the irregular sequence of burstings with overlapped spiking, we first checked how many spikes a cell is able to fire when it is released from inhibition and inter-spike intervals between these spikes. If a cell fires a series of action potentials after release from inhibition , a cell is said to have post-inhibitory rebound (PIR) burst property and STN cell has this property [28]. Duration of PIR burst, patterns of action potentials within PIR burst, and, in turn, number of spikes per burst depend on intrinsic properties of cell and network.

First, we note that *T*-type current is responsible for a PIR property of STN cell. In terms of slow variables, PIR property depends on gating variable *r* of *T*-type current, which determines the level of availability of *T*-type current. In the plane of slow variables (σ, *r*), larger value of *r* means the possibility of longer stay in spiking region when a cell is released from inhibition, hence a larger number of spikes within a burst is implicated.

To have larger *r*, sufficient level and duration of hyperpolarization are required. As was explained above, the maximal level *r* can reach under inhibition is governed by $r_\infty(V)$. If a cell is sufficiently hyperpolarized (hence the overall level of membrane potential during inhibition is sufficiently low), then $r_\infty(V)$ becomes close to one. Recall that degree of hyperpolarization is modulated by $g_{syn}$. Thus, if $g_{syn}$ is sufficiently large, then we may assume that $r_\infty(V) = 1$. In fact, this was assumed and used in the analysis of cluster solution in [20]. Over the range of $g_{syn}$ values considered in the current study, however, we have only intermediate level of $r_\infty(V)$. For example, $r_\infty^S$ was 0.6606 when $g_{syn}$ = 0.86 as we saw in previous section.

To reach the maximum level set up by $g_{syn}$, say averaged $r_\infty(V)$, a cell also needs to be inhibited sufficiently long by a sequence of action potentials with short inter-spike intervals. The requirement of sufficiently long inhibition is obvious because it takes some time for *r* to converge to the averaged $r_\infty(V)$ due to the slow time scale of the process. On the other hand, firing rate of action potentials within a PIR burst should be sufficiently high. If the inter-spike interval between inhibitory inputs is large, then cell may be depolarized and *r* begins to decrease. This may result in the lowering of overall level of *r* even though a cell gets sufficiently long hyperpolarization. For example, in the previous section the maximum level set up by $g_{syn}$ is $r_\infty^S = 0.6606$ while the level of *r* when a cell is released from inhibition is $r_*^u = 0.45$ and the following maximum value of *r* is 0.521 (Fig. 7). This may be due to the insufficient duration of inhibition (just 2 spikes) and relatively larger inter-spike intervals.

Second, decay rates of *r* and σ and their ratio also play an important role in the generation of spikes within PIR burst because trajectory of (σ, *r*) in the slow phase plane is modulated by them after release from inhibition. For example, suppose that σ decays fast and *r* decays slowly. Then, (σ, *r*) moves leftward almost horizontally in the slow phase plane. Hence it quickly passes through the low frequency band along with Σ, approaches the high frequency region and spends most of its time there. Thus we expect larger number of spikes per burst. On the other hand, if the decay rate of *r* is much faster than that of σ, then (σ, *r*) moves leftward almost vertically. In this case, (σ, *r*) would enter the spiking region, traverse the band of low frequencies and then exit without spikes or with just one spike. The 'worst' scenario is that (σ, *r*) is even not able to enter the spiking region.

Third, the number of spikes within PIR burst is also limited by self-inhibition because (σ, *r*) in the slow phase plane is pushed to the right when a cell fires. This would result in longer inter-spike intervals due to the geometry of frequency curves over the slow phase plane (Fig. 6). In addition, under current values of $g_{syn}$, self-inhibition would force the trajectory of ($\sigma$, $r$) to stay mostly in the band of low frequencies. In some cases, ($\sigma$, $r$) even has to re-enter the spiking region as seen in Figure 7. Therefore, we would expect not a large number of spikes but 2- or 3-spikes solutions with long inter-spike intervals.

Fig. 8A illustrates how the number of spikes changes with $r$ at the moment that a cell is released from inhibition. In this figure, [$Ca$] is fixed at 0.7. Recall that $\sigma$ for cell 1 is given by $g_{syn}(0.3s_1 + s_2)$ where $s_1$ is its own the synaptic variable and $s_2$ is the synaptic variable of cell 2. Upon firing of cell 2, $s_2$ approaches a value near 1 while $s_1$ is about 0.1. In this example, $\sigma_1$ approaches 0.927. Hence, we chose 0.927 as an initial condition for $\sigma$ and the corresponding vertical line in the slow phase plane is plotted at Fig. 8A. Due to the existence of low frequency region along $\Sigma$, a cell doesn't fire immediately once past $\Sigma$. As expected, number of spikes increases from 2 to 4 as $r$ increases. In each case, after the last spike, the trajectories traverse the low frequency band and leave the spiking region without additional spike.

Fig. 8B plots inter-spike intervals of trajectories shown in Fig. 8A. As before, we let $T_1$ be the time from the moment that a silent cell is released from inhibition to the point that it fires its first spike. We define $T_2$ as the first inter-spike interval, $T_3$ the second, $T_4$ the third, and so on. As implicated in Fig. 8A, $T_k(k=1,2,3,...)$ is a function of ($\sigma_0, r_0$) at the moment a cell is released from inhibition. Since $\sigma_0$ takes almost fixed values under fixed synaptic strength (for example, $\sigma \approx 0.927$ when $g_{syn} = 0.9$), $T_k(k=1,2,3,...)$ is mainly a function of $r_0$. In general, $T_k(k=1,2,3,...)$ decreases as $r_0$ increases because a trajectory in the ($\sigma$, $r$) plane, originating at larger $r_0$, gets more chances to traverse higher frequency region, that is, upper left corner of the ($\sigma$, $r$) plane. In addition, $T_k(k=1,2,3,...)$, especially $T_1$, tends to be leveled off for sufficiently large $r_0$. This happens because of the geometries of $\Sigma$ curve and spiking region, especially the low frequency band along $\Sigma$. $\Sigma$ curve is almost vertical (Fig. 8A) and the frequency curves are almost identical (Fig. 6) for sufficiently large values of $r$. Thus, trajectories in the ($\sigma$, $r$) plane, originating at sufficiently large $r_0$, pass through qualitatively similar spiking region and their corresponding $T_k(k=1,2,3,...)$ becomes similar.

We now check the effect of calcium concentration on $\Sigma$ (Fig. 8C). As calcium concentration increases from 0.4 to 0.8, $\Sigma$ moves leftward. Fig. 8C also presents values of $r$ and $\sigma$ when a cell fires its first spike after release from inhibition. These values also transition to the left along with their corresponding $\Sigma$ curves. For lower calcium levels, $\Sigma$ curves don't cross the $x$-axis. In this low [$Ca$] case, once a cell is released from inhibition, a trajectory in the ($\sigma$, $r$) plane converges to the origin and the corresponding cell activity pattern is continuous firing with very low frequency. On the other hand, $\Sigma$ crosses the y-axis for larger calcium levels as shown in Fig. 8A. In this case, a trajectory in the ($\sigma$, $r$) plane eventually escape the spiking region and cell stops firing. Now consider a trajectory that begins at ($\sigma_0, r_0$) when a cell is released from inhibition. Because the path of this trajectory is determined by ($\sigma_0, r_0$), increase of [$Ca$] does not change this trajectory but shift the spiking region to the left in the ($\sigma$, $r$) plane. As a result, this trajectory becomes closer to the low frequency band left to Σ curve and we can expect that $T_k (k=1,2,3,...)$ increases as $[Ca]$ increases (Fig. 9C). Fig. 8B also shows $T_k$ curves (dotted lines) when $[Ca]$ is not a constant but a dynamic variable. Changes in $T_k$ curves are not significant because the magnitude of $[Ca]$ fluctuations is small.

## 4.4 The 'escape' mechanism and overlapped spiking

Now we consider the mechanism underlying escape in detail using slow variables, especially $r$ and $[Ca]$. Fig. 9 demonstrates one example of escaping for $g_{syn}$ = 0.9. In this example, active cell (gray) fires three spikes and Fig. 9A shows only last two spikes. When active cell fires its last spike, silent cell (black) also fires. Fig. 9B shows the corresponding trajectories in the slow phase plane of (σ, $r$). Black dots denote time $t$ = 0ms and squares $t$ = 40ms in Fig. 9A. Small increment of σ of active cell in the middle of trajectory followed by large increment at the end of trajectory means that active cell first fired its last spike and then was inhibited by silent cell. Although active cell first fired and thus the silent cell got inhibition, silent cell was also able to fire because the membrane potential of silent cell crossed the spike threshold when it was inhibited (Fig. 9A).

We may interpret this as the competition between $T_k$ of active cell for some $k$ and $T_1$ of silent cell. Whenever active cell fires, silent cell seeks a chance to enter the spiking region. If $T_k$ of active cell becomes comparable to $T_1$ of silent cell, the possibility of escaping arises. The example shown in Fig. 2B illustrates the competition between $T_3$ of active cell and $T_1$ of silent cell, which facilitates the generation of escape. We note that a stable 3-spike bursting solution satisfies the following condition

$$T_3 < T_1 < T_4 \qquad (14)$$

that is, before active cell fires its 4$^{th}$ spike, silent cell enters the bursting region and fires. Here $T_3$ and $T_4$ are inter-spike intervals of active cell and $T_1$ is of silent cell. In the bursting solution, however, there is no distinction between $T_k$'s because the values of σ, $r$, and $[Ca]$ when silent cell is lastly released from inhibition are fixed and these values determine $T_k$'s. This relationship between $T_k$'s is satisfied for $g_{syn}$ = 0.94 and 0.96. Similarly, we have

$$T_2 < T_1 < T_3 \qquad (15)$$

for $g_{syn}$ = 0.86 and 0.88 over which stable 2-spike bursting solutions exist. In escaping case, for some reason, these relationships between $T_k$'s are disrupted due to the comparability between $T_1$ and $T_3$.

Fig. 9C shows curves of $T_1$ and $T_3$ when $g_{syn} = 0.9$ and $[Ca] = 0.66$ (black) and 0.7 (gray). Here $[Ca]$ is not a constant but a variable and the above values are initial values of $[Ca]$ when silent cell is lastly released from inhibition. We define $r_f$ to be the value of $r_0$ such that if $r_0 = r_f$, then $T_1 = T_3$ and if $r_0 < r_f$ ($r_0 > r_f$), then $T_1 < T_3$ ($T_1 > T_3$). Naturally, $r_f$ is a function of $[Ca]$ and Fig. 9C demonstrates that $r_f$ exists $[Ca] = 0.66$ and 0.7. When $g_{syn} = 0.9$, numerically computed $r_f$ is 0.474 (0.476, 0.480, 0.483, resp.) when $[Ca]$ is 0.64 (0.66, 0.68, 0.70, resp.). $r_f$ increases as $[Ca]$ increases.

Figure 10 shows $r_f$ values (middle gray lines with star marks) over a range of $[Ca]$ and $g_{syn}$ values. Except for smaller values of $g_{syn}$, $r_f$ show monotonic increase as $[Ca]$ increases. These $r_f$ values form a band from the upper-left corner to the lower-right corner, which separates 2-spike and 3-spike regular bursting solutions. Figure 10 also shows numerically calculated $r_*$ values of 2-spike (lower left four circles) and 3-spike (upper right four squares) regular bursting solutions. These two groups of $r_f$ values are extrapolated respectively using quadratic polynomials to the $g_{syn}$ values of interest, from 0.9 to 0.92. The resulting extrapolated $r_*$ values are also plotted in the same figure.

Figure 10 demonstrates the proximity between extrapolated $r_*$ values and $r_f$ values for $g_{syn}$ from 0.9 to 0.92, over which escaping activity patterns are prominent. Especially the extrapolated values of $r_*$ fall within the band when $g_{syn}$ is 0.91. We cannot use this result directly because $r_f$ values were obtained from the comparison between inter-spike intervals of the same cell while we need to compare $T_3$ of active cell and $T_1$ of silent cell. However, the closeness of $r_f$ and $r_*$ may suggest a heuristic explanation on the mechanism underlying escaping. A trajectory starting at some initial conditions near those of 2-spike bursting solution tends to approach the regular bursting solution after some time. While doing so, due to the proximity between $r_*$ and $r_f$, $T_1$ of silent cell becomes comparable to $T_3$ of active cell, which renders the possibility of escaping. Once escaping happens, both cells fall into 3-spike bursting regime. However, due to the instability of 3-spike bursting solution, both cells are pushed into 2-spike bursting solution regime again and this process repeated.

Now we look at escaping more closely in the $(r, [Ca])$ plane. Figure 11A shows the trajectories of active cell and silent cell from the time when the active cell is released from inhibition to escaping in the $(r, [Ca])$ plane. Triangles denote starting points of trajectories. At the first star, silent cell fires its last spike and active cell is released from inhibition. For the active cell, these values of $r$ and $[Ca]$ determine inter-spike intervals, which will be compared to $T_1$ of silent cell. Whenever the active cell fires, $T_1$ is determined by $r$ and $[Ca]$ values of the active cell at that moment. At the moment of the second spike of the active cell, the trajectory of the silent cell lies close to the open square, the extrapolated 2-spike bursting solution point. As shown in this figure, the trajectories tend to approach the extrapolated 2-spike bursting solution point. Figure 11B illustrates how the closeness of the square to the right vertical line causes escaping. As expected in Fig. 11A, $T_3$ value of the active cell is close to $T_1$ value of silent cell, which facilitates escaping.

In summary, escaping may result from the closeness between $r_*$ and $r_f$. In $(r, [Ca])$ plane, each trajectory approaches the extrapolated values of $r$ and $[Ca]$ in regular 2-spike bursting solution over time. Due to the proximity of $r_*$ and $r_f$, $T_1$ becomes comparable to $T_3$ and this provides good ingredient for escaping. Once escaping occurs, σ values of both cells reach larger values due to the combinations of self-inhibition and inhibition from other cell (Fig. 9B). Because of this increased amount of σ, it requires longer time for the escaping cell to enter the spiking region to fire an additional spike. This extended depolarizing period results in a larger value of $r$ and may enable the escaping cell to fire additional two spikes depending on the availability of *T*-type current. If the escaping cell fires additional two spikes, then the other cell would fire three spikes when it is released due to the extended duration of inhibition. Hence we have 3-spike alternating bursting patterns. In $(r, [Ca])$ plane, this means that $(r, [Ca])$ of silent cell when it is released from inhibition lies on the right side of $r_f$ line where $T_1 > T_3$. But after some time, due to the insufficient hyperpolarization, trajectories of two cells come back into 2-spike region and this process repeats.

While the results in this section may provide a scenario for switching between 2- and 3- spike burstings with escaping, as stated in Section 4.1, numerical study shows that the slow fluctuation of $[Ca]$ appears to be responsible for the irregularity of the generation of escaping. Because the possibility of escaping is very sensitive to the relative timing of the suppressed and active cells, small differences of $[Ca]$ may determine whether the suppressed cell can escape or not. Further studies may be needed to explain how the slow fluctuations of $[Ca]$ enhance the robustness of the irregularity of the activity patterns over the synaptic strength.

## 5. Discussion

In the present work, we considered a reciprocally connected inhibitory network with self-inhibition consisting of two cells. There are intervals of coupling parameters over which activity patterns exhibit an irregular bursting sequence. This characteristic irregular bursting sequence may be attributed to the frequent occurrence of overlapped spiking ('escape') due to the competition of the two cells. We show that the irregularity of overlapping and the resulting variability in the number of spikes within a burst significantly reduces the degree of a phase-locking between the two cells. In this case, the fine temporal structure of phase-locking between two cells is characterized by intermittent occurrences of relatively short deviations from the phase-locked state, which is similar to what was observed in experimental data in the brains of parkinsonian patients [13].

To investigate the mechanism underlying this intermittent occurrence of escaping and the resulting bursting patterns, we considered the dynamics of slow variables and used geometric methods for analysis. For the fast/slow analysis, we chose three slow variables, the gating variable $r$ of $T$-type current, the total synaptic input $\sigma$, and calcium concentration $[Ca]$. We constructed a two-dimensional bifurcation diagram using $r$ and $\sigma$ for constant $[Ca]$ to find the $\Sigma$-curve, which divides the slow phase plane into silent and sustained spiking regions. Slow variation of $[Ca]$ moves this curve horizontally. The sustained spiking region is roughly divided into two sub-regions according to frequencies, a low frequency band along $\Sigma$ and a high frequency region. Since the activity patterns of the cells can be qualitatively described on the slow phase plane ($\sigma$, $r$) along the trajectory of the system, the geometry of the slow phase plane including the $\Sigma$-curve is an important determiner of the dynamics.

Our results demonstrate that an irregular sequence of burstings results from the complex interplay between network architecture, decay rates of the synaptic variable and the gating variable of $T$- type current, and other intrinsic properties of cells which are responsible for the geometry of the slow phase plane, especially structure of $\Sigma$ and low frequency band along it. Overlapped spiking, the main cause of the irregular sequence of burstings, is the result of competition between two cells; the suppressed cell may enter the bursting region while the active cell fires additional spike. In more technical terms, in the system considered in this study, this overlapped spiking happens due to the competition between $T_1$ of silent cell and $T_3$ of active cell, where the former is the time from the moment when the suppressed cell is released from inhibition to its first spike and the latter is the time between second and third spikes. Over the stable bursting solution regime, the value of $r$ when a cell is released from inhibition, say $r_0$, converges to a stable fixed point $r_*$ (Eq. (5)), and there is a fixed relationship between interspike intervals. For example, $T_2 < T_1 < T_3$ for 2-spike solution and $T_3 < T_1 < T_4$ for 3-spike solution. As $g_{syn}$ increases from 2-spike bursting regime to 3-spike bursting regime, $r_*$ increases to approach $r_f$ where $T_1$ equals $T_3$. Due to this proximity, $r_0$ approaches $r_f$ during the repeated burstings. As a result, $T_1$ of silent cell and $T_3$ of active cell become comparable. In the 2-spike bursting region, this proximity also allows a cell to fire an additional spike so that the other cell may be pushed into the 3-spike bursting region. In summary, the closeness of $r_*$'s and $r_f$ acts as a driving force which pushes the solution from 3-spike bursting region into 2-spike region and vice versa. While the switching between 2- and 3-spike burstings with escaping results from the proximity of $r_*$ and $r_f$, the robustness of irregularity in overlapping of spikes seems to come from the slow fluctuation of $[Ca]$. This may be due to the fact that the possibility of escaping depends on the small difference between the phases of two cells. Although the amount of change of $[Ca]$ is small, its effect is significant for the described activity patterns to be robust as to the synaptic strength, the dopamine-dependent variable.

This proximity of $r_*$ and $r_f$ depends on several factors. An appropriate level of the synaptic coupling strength $g_{syn}$ pushes $r$ high enough under inhibition. Comparable decay rates of the synaptic variable $\sigma$ and the gating variable $r$ along with the self-inhibition architecture and the bending part of Σ let the system stay mostly in the low frequency region in the plane of the slow variables (σ, r).

Current study is partially based on results of previous studies of STN-GPe network [20,29], in which sufficiently strong or weak coupling strengths made a rather complete mathematical analysis possible. In the current study,the intermediate level of synaptic strength along with self-coupling network architecture makes activity patterns more realistic, however the dynamics and its analysis is much more complex. It would be desirable to obtain one or two dimensional maps to explain the existence of chaotic behavior or the route to chaos as some physiologically relevant parameters change as in [20,29]. However, the chaotic dynamics observed in this study shows sensitivity to the small difference between values of slow variables, hence analysis using averaged equations doesn't work. To trace the trajectories because is challenging due to the following limitations: intermediate level of synaptic strength, the geometry and low frequency band of Σ curve, and network architecture with self-inhibition. Instead, the present study outlines a scenario for a realistically complex dynamics, advancing us beyond the extreme cases of weak and strong coupling studied earlier.

The considered mechanism of generation of intermittent synchronous patterns requires some tuning of the parameters of neurons and synapses, which leads to the appropriate timing of spikes. Thus the question of the overall robustness of this mechanism remains open. However, in numerical experiments this tuning does not appear to be very fine and allows for some variation of the parameters characterizing the timings of synapses and ionic channels. Moreover, real living neuronal networks with larger numbers of neurons and some degree of inhomogeneity may loosen the requirements for tuning as the variation of parameters in the network may make the necessary similarity of the time-scales easier to achieve. In addition, robustness to noise further supports the robustness and generic nature of the considered phenomenon.

This suggests that the considered mechanism is relevant to the experimentally observed cases of intermittent dynamics. As we described in the Introduction, transient nature of dynamics in neural systems may be quite general in the nervous system [16,17]. In particular the networks of the basal ganglia, which inspired the model considered here, exhibit intermittent phase locking, similar to that considered here [13]. Our recent modeling studies of larger and more realistic networks showed that the parameter region for realistic intermittent synchrony is located in the boundary between synchronized (presumably extreme pathological) and nonsynchronized (presumably normal) dynamics. It was conjectured there that closeness to the synchrony regime would yield easy formation and dissociation of synchronized neuronal assemblies [1]. The present study provides one possible generic mechanism for irregular escaping events which could be responsible for intermittent disruptions of phase-locking between two cells.

This modeling also gives consistent results in that dopamine-dependent synaptic coupling strength $g_{syn}$ controls various activity patterns and the transitions between them. As we showed above, an intermediate level of synaptic strength is crucial in the generation of overlapped spiking in conjunction with the geometry of Σ. Under the loss of dopamine, STN cell becomes sensitive to the inputs from GPe [22,23,24,25]. Hence the loss of dopaminergic cells in Parkinson's disease may result in an increase of sensitivity of the STN cell and of $g_{syn}$. These experimental results are consistent with our current study: up to some level of $g_{syn}$ the network exhibits low-correlation activity patterns which correspond to the normal state of the basal ganglia. But for intermediate value of $g_{syn}$ the network shows more synchronized activity, which is interrupted by desynchronized events such as overlapped spiking. This fits very well with the comparison of the larger model network with experimental data [1], which suggests that parkinsonian state is characterized by intermediate (rather than large) values of synaptic strength $g_{syn}$ and other related parameters.

The mutual inhibition of bursting neurons (direct or through excitatory cells) in our network may be a fairly generic cellular and network set-up. Thus we think that the considered mechanisms of intermittent phase-locking have implications for neural systems beyond the basal ganglia, including subcortical areas or, possibly, central pattern generators.

## Acknowledgements

We thank Dr. R.M. Worth for his comments on the manuscript.
This study was supported by NIH grant R01NS067200 (NSF/NIH CRCNS program).

# Figures

**Figure1**. Network architecture. (A) Reciprocally connected large STN-GPe network. Arrow indicates excitatory connection and circle indicates inhibitory connection. (B) Reduced network of two inhibitory cells with self-inhibition.

**Figure 2.** (A) Averaged coherence over 0-100 Hz band (black) and 10-30 Hz band (gray) between voltages of two cells. Coupling strength $g_{syn}$ varies from 0.1 to 2 with 0.1 stepsize. (B) Phase-locking index $\gamma$ in dependence on $g_{syn}$. Different lines denote different strength of self-coupling; default value is 30% (black solid) and two other strengths, 25% (dotted) and 35% (gray) are also shown for comparison. (C) Activity patterns with irregular sequence of burstings with overlapped spikes when $g_{syn}= 0.9$.

**Figure 3.** Return map $r_{n+1}$ vs. $r_n$ for different values of the synaptic strength parameter $g_{syn}$ (A) presents five different values of $g_{syn}$ : 0.86(square), 0.88(circle), 0.94(triangle), 0.96(diamond) and 0.92(gray dots). (B) More complex dynamics observed for $g_{syn} = 0.9$. (C) Maximal Lyapunov exponents (MLEs) over a range of refined $g_{syn}$ values between 0.9 and 0.92. When $g_{syn}$ lies between 0.9 and 0.91, MLE values are significant as compared to other range of $g_{syn}$ values. Over this range of $g_{syn}$ values, chaotic activity patterns may be robust.

**Figure 4.** Irregular partially-synchronous dynamics of reduced model when $g_{syn} = 0.9$. (A) An example of escaping. Upper panel shows voltage profiles and lower panel shows corresponding phases for two neurons. (B) Return map for phases $\phi_{i+1}$ vs. $\phi_i$. (C) Histrogram of durations of desynchronization events. Black bars come from experimental data [13].

**Figure 5.** (A) STN cell activity patterns when $g_{syn} = 0.9$. Upper panel shows voltage profile and lower panel shows slow variables $r$ (black solid), $\sigma$ (lower gray solid), and $[Ca]$ (black dotted). Refer to Eqs. (2)-(4). The time course of $[Ca]$ is also plotted with the scale on the right vertical axis (upper gray solid). (B) Bifurcation diagram of the fast subsystem for $r = 1$ and $[Ca] = 0.7$. There is a saddle-node bifurcation point (SN) at $\sigma = \sigma_{SN}$ and also a subcritical Hopf bifurcation point on the upper branch where an unstable periodic orbit begins to be turning to a stable periodic orbit. This stable periodic orbit becomes a saddle-node homoclinic orbit when $\sigma = \sigma_{SN}$. Stable (unstable) fixed points and limit cycles are in thick black (thin gray).

**Figure 6.** The frequency of firing in dependence on the slow variables σ and *r*. (A) Σ-curve (gray line in the (σ, *r*) plane) divides the space of the slow variables (σ, *r*) into silent and sustained spiking regions. Over the sustained spiking region, the curves of frequencies are plotted for various *r* (0.25, 0.3, 0.35, 0.4, 0.5, 0.6, 0.7, 0.8, 0.9 and 1, left to right). (B) Another view of frequency curves from the part (A). For larger values of *r*, the frequency of the periodic solution decreases almost linearly as σ decreases and then sharply decreases near Σ. For *r* ≥ 0.5, the frequency curves are almost identical.

**Figure 7.** Two-parameter bifurcation diagram with projection of 2-spike out-of-phase bursting solution. The close-to-vertical curve in the middle of the figure is the Σ-curve shown in Fig. 6 when [*Ca*] = 0.7. The moment when active cell fires its last spike is denoted by lower circle and upper circle denotes (σ, *r*) of silent cell at that moment. Open circle denotes the moment that silent cell is released from inhibition. The moment when the silent cell fires its first spike is denoted by upper square and lower square denotes (σ, *r*) of active cell at that moment. $T_1$ is the time needed for silent cell to fire its first spike after it is released from inhibition (from open circle to upper square) and $T_2$ is inter-spike interval between first and second spike. Black line is the trajectory of active cell from its first spike until the moment that it gets inhibition to leave the spiking region. The counterpart of silent cell is denoted by gray trajectory.

**Figure 8.** Spike firing for different values of *r* and [*Ca*]. (A) The number of spikes that a cell fires depends on the level of *r* when the cell is released from inhibition. The vertical line (σ = 0.927) is divided into three intervals according to number of spikes, 2, 3, and 4. Boundaries between these intervals are marked by two dots (*r* = 0.45 and 0.58). Three exemplary cases for each number of spikes are presented for *r* is 0.4 (2 spikes, dotted), 0.5 (3 spikes, black solid), and 0.6 (4 spikes, gray solid). (B) Plot of $T_1$ (left black solid), $T_2$ (left gray solid), $T_3$ (right gray solid) and $T_4$ (right black solid) as a function of *r*. The dotted lines are $T_k$ curves for $k = 1,2,3$ and 4 when [*Ca*] is slow variable. (C) Σ-curves for several [*Ca*] levels from 0.4 to 0.8 with stepsize 0.1 (left to right). Values of *r* and σ are also plotted when a cell fires its first spike for various initial values of *r*, from 0.35 to 0.7 with the stepsize 0.05 (from bottom to top). Synaptic strength $g_{syn} = 0.9$.

**Figure 9.** An example of 'escaping' for $g_{syn} = 0.9$; one cell fires three spikes and during the firing of its last spike, the other cell also fires. (A) Voltage traces of both cells (only last two spikes are shown in this figure, black line is escaping cell). (B) The corresponding trajectory in the phase space of slow variables. Black dots (squares) correspond to time *t* = 0 (*t* = 40) in part (A). (C) $T_1$ (solid) and $T_3$ (dotted) curves as a function of r when [*Ca*] is variable. Initial conditions for [*Ca*] are 0.66 (black) and 0.7 (gray).

**Figure 10.** $r_*$ and $r_f$ values for the range of $g_{syn}$ and different values of $[Ca]$. The lower four circles and upper four squares are numerically computed $r_*$'s for 2-spike and 3-spike regular bursting solutions. The middle three circles and squares are obtained by quadratic extrapolation. Numerically computed $r_f$'s for different $[Ca]$ levels are also shown (gray stars and lines, $[Ca]$ level increases from 0.6, bottom on the right, to 0.7, top on the right, with the stepsize of 0.02).

**Figure 11.** Escaping revisited. (A) An escaping example in the $(r, [Ca])$ plane. Close to vertical lines divide the plane according to the number of spikes when a cell is released from inhibition (Fig. 8). Left to the dotted line is 1-spike region and right 2-spike region. Similarly, right to the dashed line is 3-spike region. Solid line is the line of $r_f$ in the 3-spike region. Four circles between middle and right vertical lines denote the values of $r$ and $[Ca]$ when silent cell is lastly inhibited in a regular 2-spike bursting solution. From bottom to top, $g_{syn}$ values are from 0.86 to 0.89. Square at the top of this array of circles denotes the extrapolated values of $r$ and $[Ca]$ when $g_{syn}$ is 0.9. Lower trajectory is active cell and upper trajectory is silent cell and triangles denote starting points of each trajectory. Stars denote the moments when cells get inhibition either from itself or from other cell. (B) Three gray solid lines are $T_k$ curves using the calcium level when the active cell is lastly inhibited. Dots on these curves denote $T_2$ and $T_3$ of active cell computed from the value of $r$ when the active cell is released. Black curve is $T_1$ of silent cell when it gets second inhibition and square denotes actual $T_1$ value at that moment.

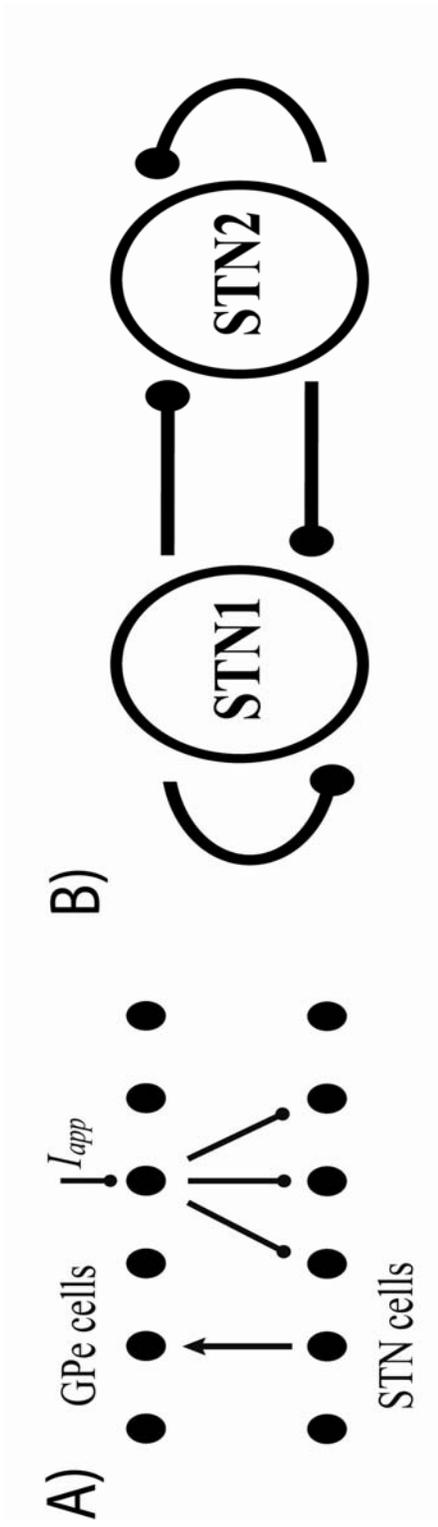

Figure 1

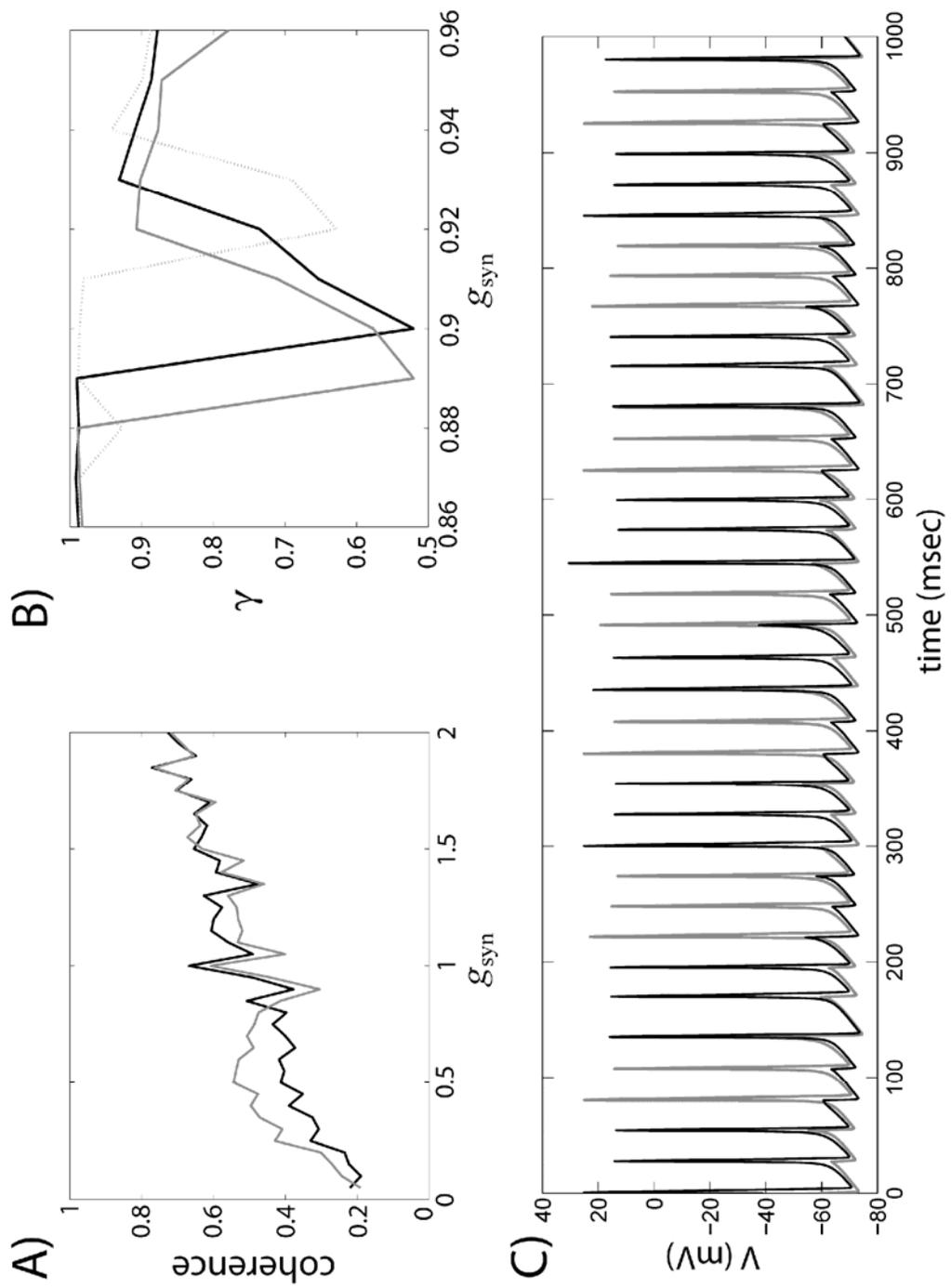

Figure 2

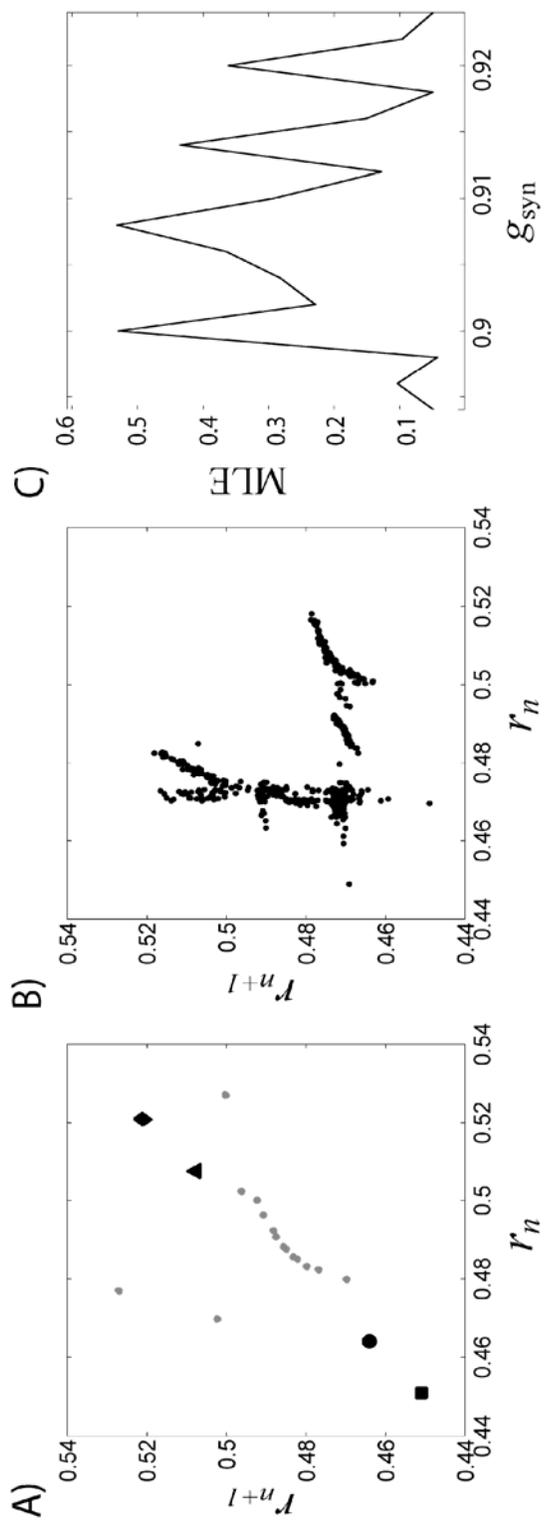

Figure 3

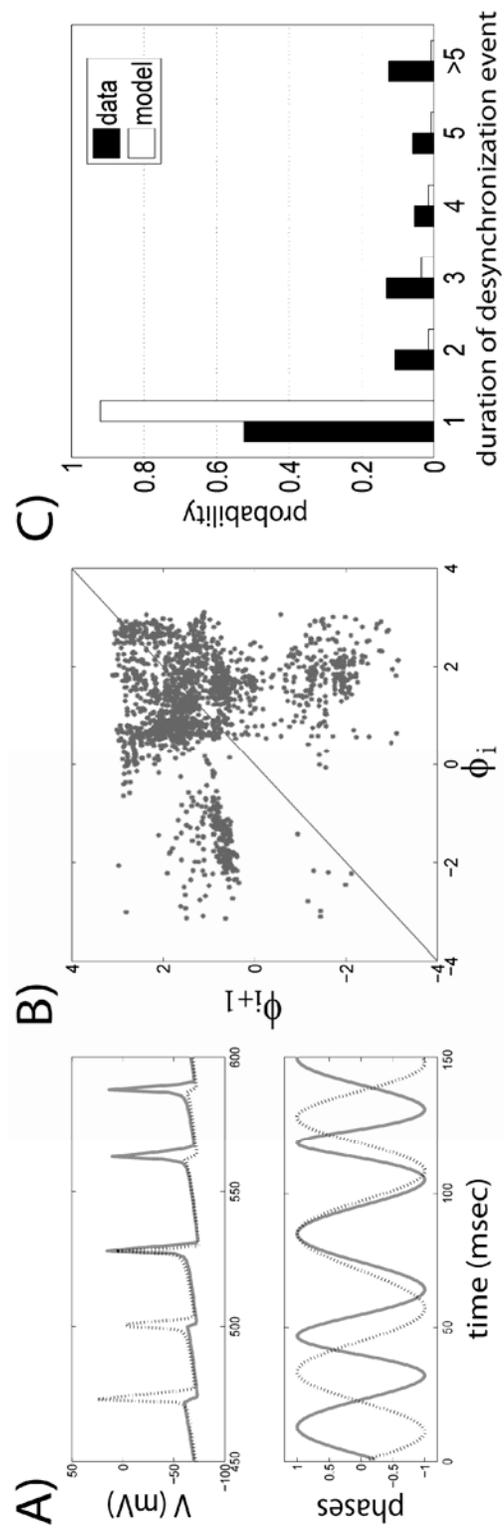

Figure 4

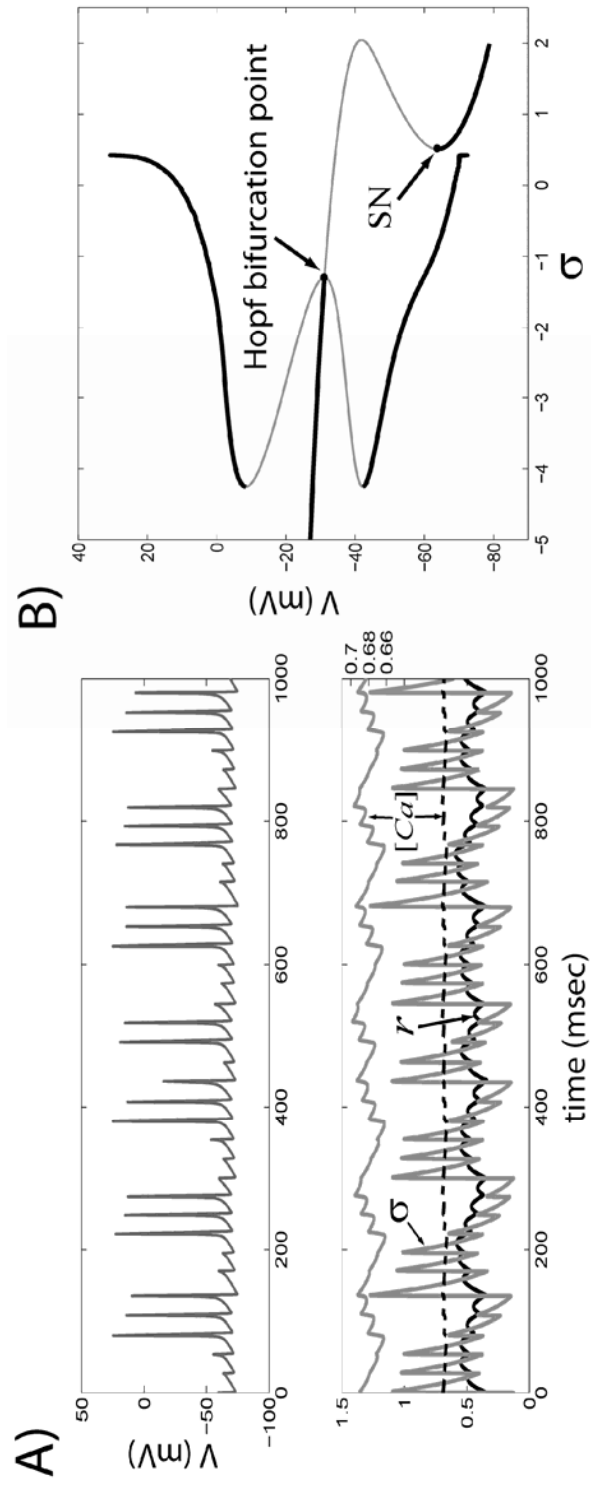

Figure 5

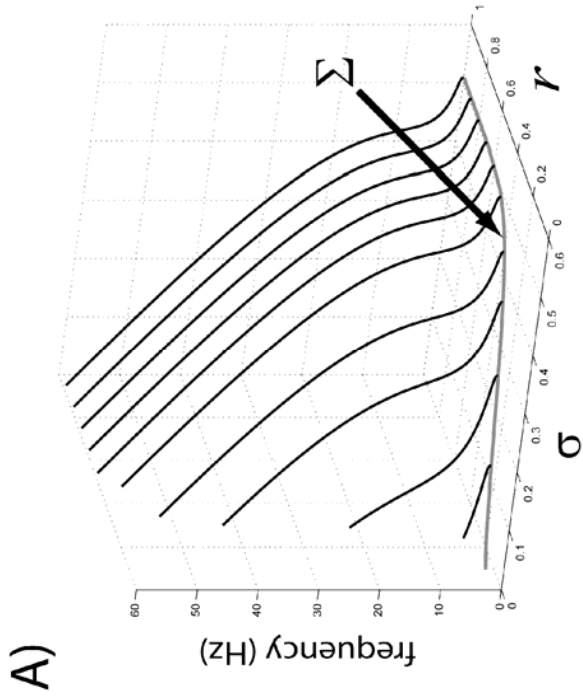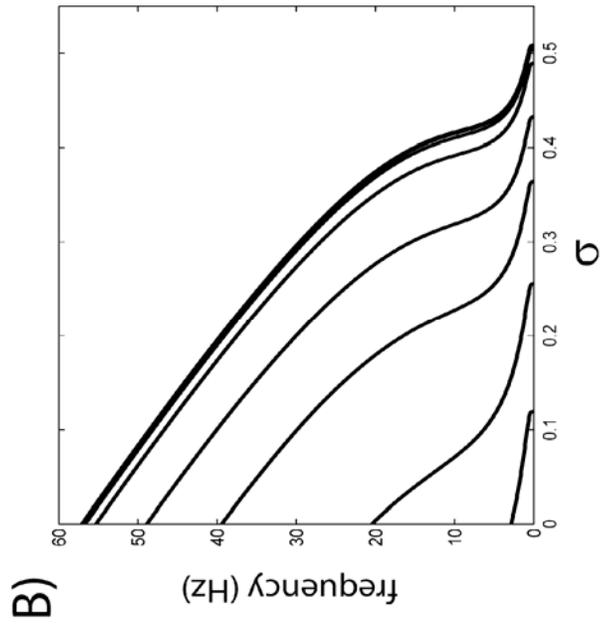

Figure 6

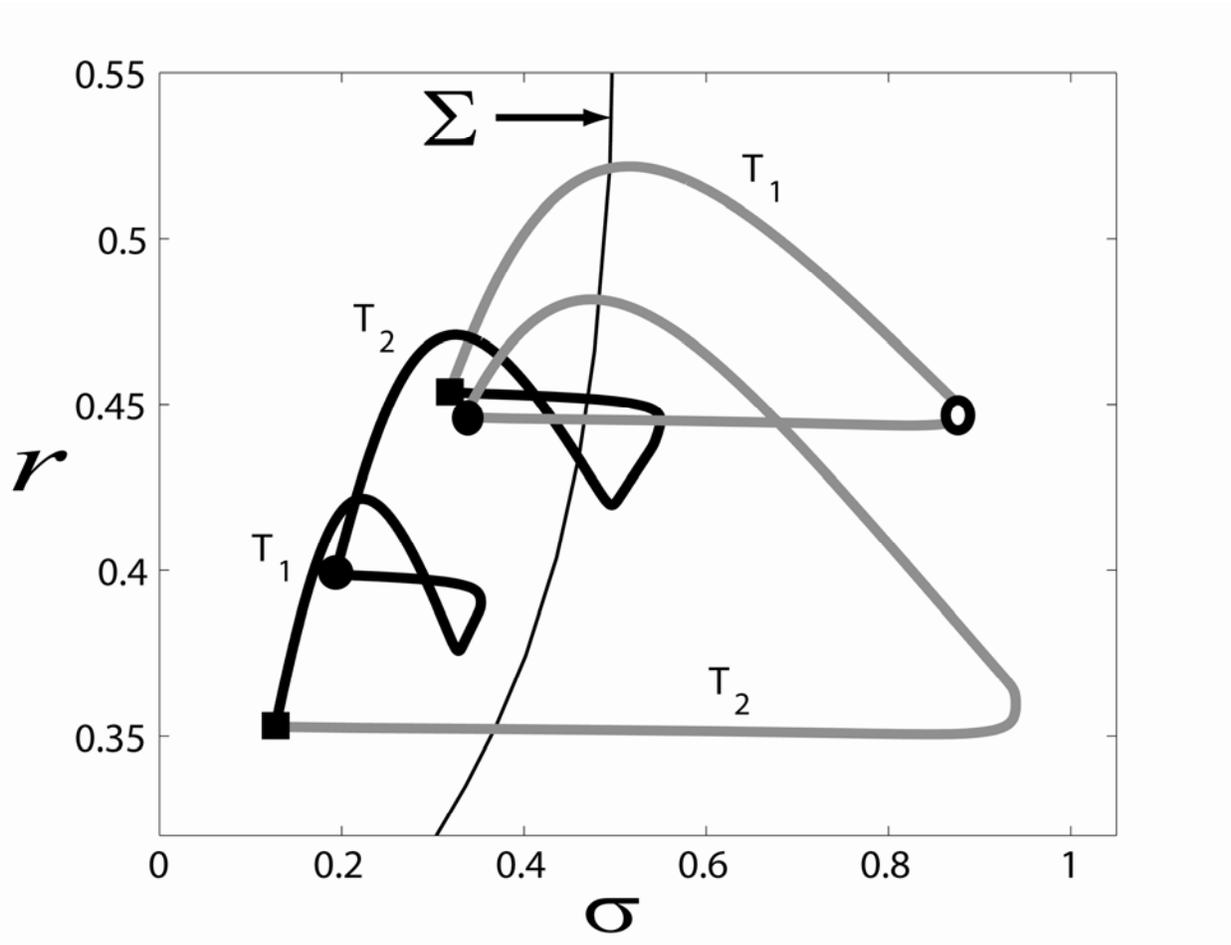

Figure 7

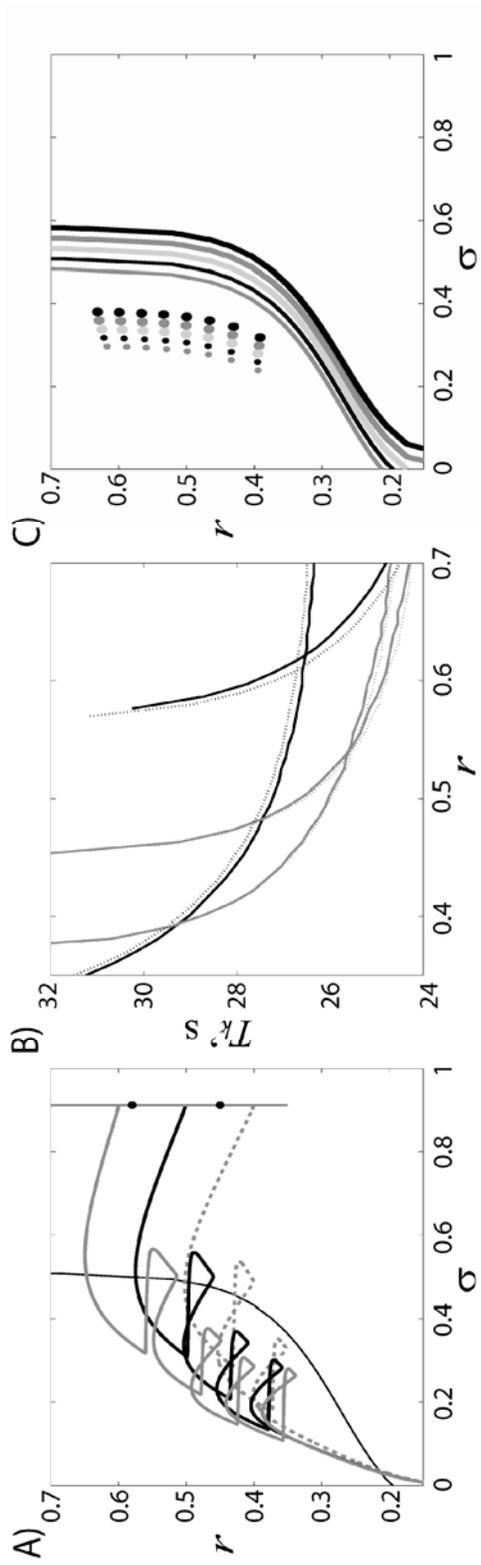

Figure 8

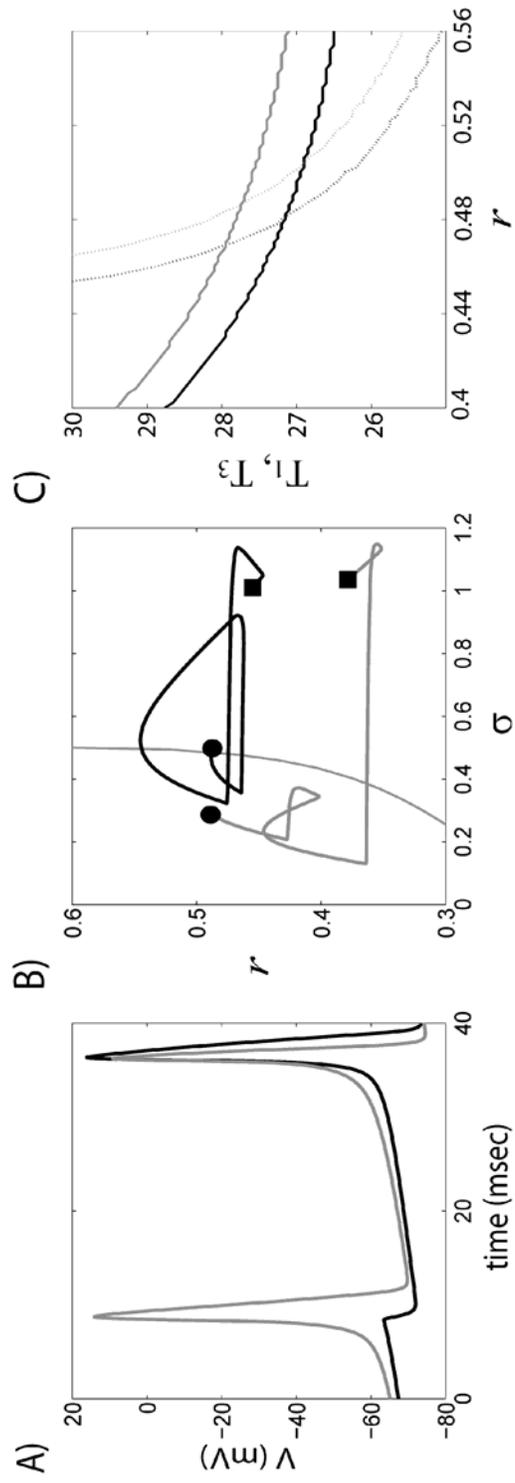

Figure 9

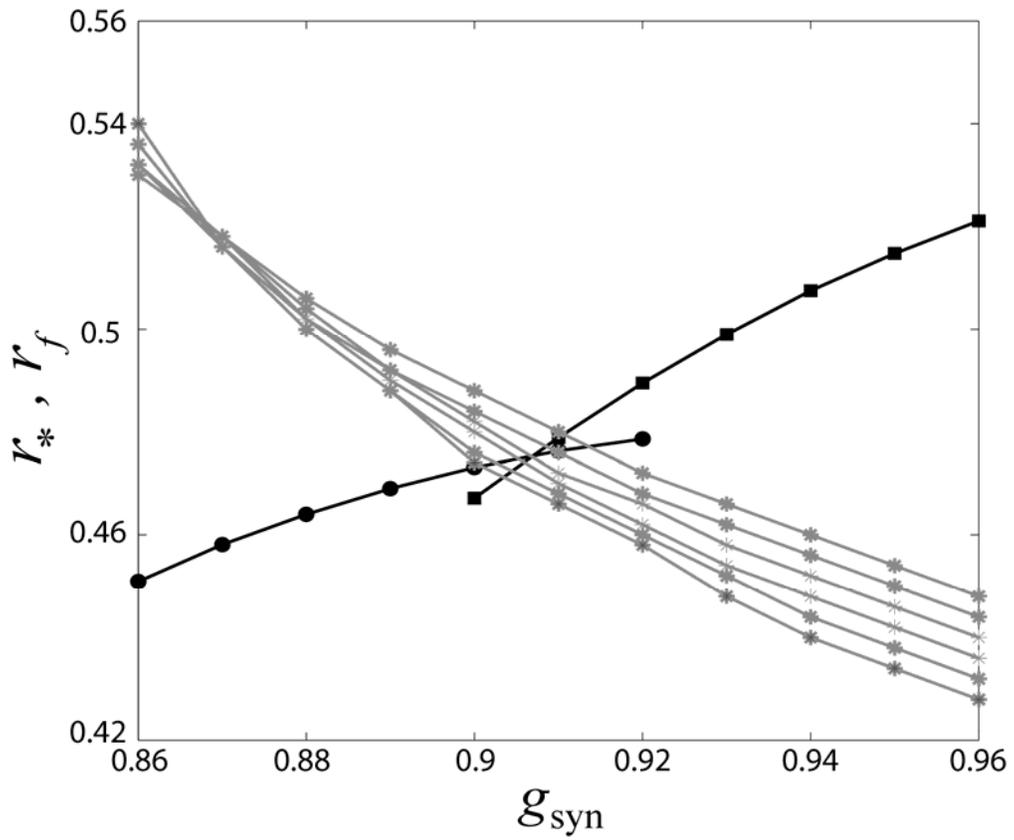

Figure 10

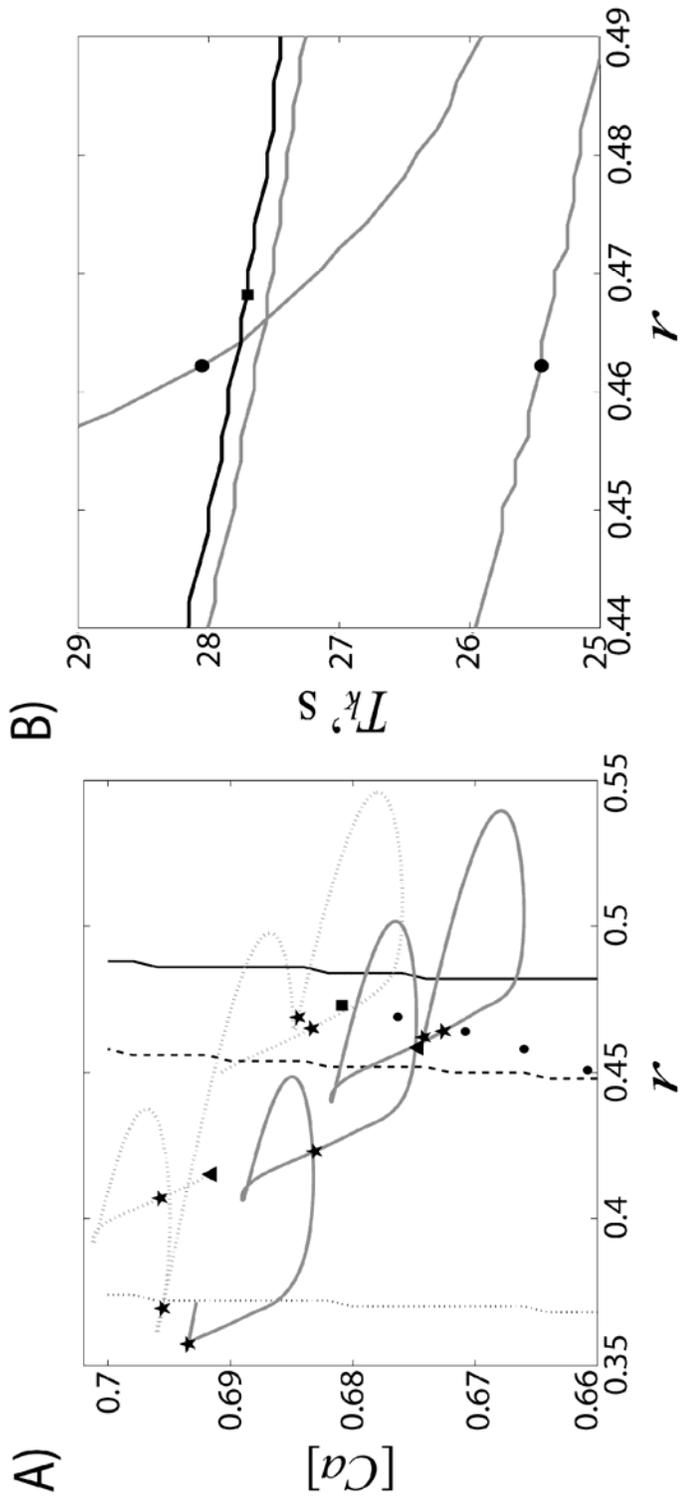

Figure 11